 \documentclass[structabstract]{aa}  
\usepackage{graphicx}
\usepackage{txfonts}
\usepackage{natbib}
\bibpunct{(}{)}{;}{a}{}{,}

\begin{document}

   \title{ The orbits of subdwarf B + main-sequence binaries}

   \subtitle{I: The sdB+G0 system PG\,1104+243}

   \author{J.~Vos
          \inst{1}
          \and
	  R.H.~\O{}stensen
	  \inst{1}
	  \and
	  P.~Degroote
	  \inst{1}
	  \and
	  K.~De~Smedt
	  \inst{1}
	  \and
	  E.M.~Green
	  \inst{2}
	  \and
	  U.~Heber
	  \inst{3}
	  \and
	  H.~Van~Winckel
	  \inst{1}
	  \and
	  B.~Acke
	  \inst{1}
	  \and
	  S.~Bloemen
	  \inst{1}
	  \and
	  P.~De~Cat
	  \inst{4}
	  \and
	  K.~Exter
	  \inst{1}
	  \and
	  P.~Lampens
	  \inst{4}
	  \and
	  R.~Lombaert
	  \inst{1}
	  \and
	  T.~Masseron
	  \inst{5}
	  \and
	  J.~Menu
	  \inst{1}
	  \and
	  P.~Neyskens
	  \inst{5}
	  \and
	  G.~Raskin
	  \inst{1}
	  \and
	  E.~Ringat
	  \inst{6}
	  \and
	  T.~Rauch
	  \inst{6}
	  \and
	  K.~Smolders
	  \inst{1}
	  \and
	  A.~Tkachenko
	  \inst{1}
          }


   \institute{
    Instituut voor Sterrenkunde, KU Leuven, Celestijnenlaan 200D, B-3001 Leuven, Belgium\\
    \email{jorisv@ster.kuleuven.be}
    \and
    Steward Observatory, University of Arizona, 933 North Cherry Avenue, Tucson, AZ 85721, USA
    \and
    Dr. Remeis-Sternwarte Astronomisches Institut, Universit\"{a}t Erlangen-N\"{u}rnberg, D-96049 Bamberg, Germany
    \and
    Royal Observatory of Belgium, Ringlaan 3, B-1180 Brussels, Belgium
    \and
    Universit\'{e} Libre de Bruxelles, C.P. 226, Boulevard du Triomphe, B-1050 Bruxelles, Belgium
    \and
    Institut f\"{u}r Astronomie und Astrophysik, Universit\"{a}t T\"{u}bingen, Sand 1, D-72076 T\"{u}bingen, Germany
             }

   \date{Received \today; accepted ???}

 
  \abstract
   {The predicted orbital period histogram of a subdwarf B (sdB) population is bimodal with a peak at short ( $<$ 10 days) and long ( $>$ 250 days) periods. Observationally, however, there are many short-period sdB systems known, but only very few long-period sdB binaries are identified. As these predictions are based on poorly understood binary interaction processes, it is of prime importance to confront the predictions to well constrained observational data. We therefore initiated a monitoring program to find and characterize long-period sdB stars. }
   {In this contribution we aim to determine the absolute dimensions of the long-period binary system PG\,1104+243 consisting of an sdB and a main-sequence (MS) component, and determine its evolution history.}
   {High-resolution spectroscopy time-series were obtained with HERMES at the Mercator telescope at La Palma, and analyzed to determine the radial velocities of both the sdB and MS components. Photometry from the literature was used to construct the spectral energy distribution (SED) of the binary. Atmosphere models were used to fit this SED and determine the surface gravity and temperature of both components. The gravitational redshift provided an independent confirmation of the surface gravity of the sdB component.}
   {An orbital period of 753 $\pm$ 3 d and a mass ratio of $q$ = 0.637 $\pm$ 0.015 were found for PG\,1104+243 from the radial velocity curves. The sdB component has an effective temperature of $T_{\rm{eff}}$ = 33500 $\pm$ 1200 K and a surface gravity of $\log{g}$ = 5.84 $\pm$ 0.08 dex, while the cool companion is found to be a G-type star with $T_{\rm{eff}}$ = 5930 $\pm$ 160 K and $\log{g}$ = 4.29 $\pm$ 0.05 dex. When a canonical mass of $M_{\rm{sdB}}$ = 0.47 $M_{\odot}$ is assumed, the MS component has a mass of $M_{\rm{MS}}$ = 0.74 $\pm$ 0.07 $M_{\odot}$, and its temperature corresponds to what is expected for a terminal age main-sequence star with sub-solar metalicity.}
   {PG\,1104+243 is the first long-period sdB binary in which accurate and consistent physical parameters of both components could be determined, and the first sdB binary in which the gravitational redshift is measured. Furthermore, PG\,1104+243 is the first sdB+MS system that shows consistent evidence for being formed through stable Roche-lobe overflow. An analysis of a larger sample of long-period sdB binaries will allow for the refinement of several essential parameters in the current formation channels.}

   \keywords{stars: evolution -- stars: fundamental parameters -- stars: subdwarfs -- stars: binaries: spectroscopic}

   \maketitle
%

\section{Introduction}\label{s-intro}
Subdwarf B (sdB) stars are considered to be core helium burning stars with a very thin (M$_{\rm{H}}$ $<$ 0.02 $M_{\odot}$) hydrogen envelope, and a mass close to the core helium flash mass of 0.47 $M_{\odot}$. Their existence provides stellar evolutionary theory with a challenge: how can a red giant lose its entire hydrogen envelope before starting core helium fusion? \citet{Mengel75} was the first to propose the hypothesis of binary evolution to form sdB stars. Currently there are five evolutionary channels that are thought to produce sdB stars, all of which are binary evolution channels where interaction physics play a major role \citep{Heber09}. The first two are common envelope (CE) ejection channels. In the first of these, a CE is formed by unstable mass transfer and expelled due to energy transfer from the shrinking orbit to the envelope.
In the second CE channel, the envelope is not ejected in the initial dynamical phase, but at a later stage, when the companion penetrates the initial radiative layer of the giant. These channels produce short-period sdB binaries with $P_{\rm{orb}}$ = 0.1 -- 10 d. Furthermore, there are two stable Roche-lobe overflow (RLOF) channels. In the first, distinct mass transfer starts near the tip of the first giant branch, and in the second, mass transfer occurs when the binary is passing through the Hertzsprung gap. Subdwarf B binaries formed in this way are predicted to have longer periods ranging from 10 to over 500 days. All of the above scenarios result in a very small mass range for the sdB component, peaking at $M_{\rm{sdB}}$ = 0.47 $\pm$ 0.05 $M_{\odot}$. The last channel that can produce an sdB star is the double white dwarf merger channel, where a pair of close helium-core white dwarfs loses orbital energy through gravitational waves, resulting in a merger. This channel forms a single subdwarf star with a higher mass, up to 0.65 $M_{\odot}$ \citep{Webbink84}. 

The binary interaction processes that form these sdB binaries are very complex, and many aspects are still not completely understood. Formally, they are described using several parameters such as efficiency of envelope ejection, accretion efficiency, physical interaction during common envelope phase, etc. Most of these are currently poorly constrained by observations. The most common tool to constrain evolution paths is by performing a binary population synthesis (BPS) and comparing the resulting population to an observed population for different values of these parameters. \citet{Han02, Han03}, \citet{Nelemans10} and \citet{Clausen12} have performed such BPS studies. The main results are that subdwarf progenitors with a low-mass companion will have unstable mass transfer on the RGB, during which their orbit will shrink, resulting in short-period sdB binaries. Subdwarf progenitors with a more massive companion will have expanded orbits, resulting in long-period sdB binaries.  Several observational studies \citep{Maxted01,Morales03, Copperwheat11} have focused on short-period binaries, and found that about half of all sdB stars reside in short-period (P$_{\rm{orb}}$ $<$ 10 d) binaries. Longer period sdB binaries have been observed \citep[e.g.][]{Green01}, but no definite orbits have been established. 

\citet{Clausen12} concluded that, although the currently known population of sdB binaries with a white dwarf (WD) or M dwarf companion can constrain some parameters used in the binary evolution codes, many parameter remain open. Furthermore they concluded that sdB binaries with a main-sequence (MS) component will be able to provide a better understanding of the limiting mass ratio for dynamically stable mass transfer during the RLOF phase, the way that mass is lost to space, and the transfer of orbital energy to the envelope during a common envelope phase. The period distribution of sdB + MS binaries can support the existence of stable mass transfer on the RGB. If long-period sdB + MS binaries are found, they are thought to be formed during a stable mass transfer phase in the RLOF evolutionary channel. If no long-period sdB + MS binaries can be found, this would indicate that mass transfer on the RGB is unstable.

The goal of this paper is to describe the methods we used to determine radial velocities of both the main-sequence and subdwarf B component in sdB + MS systems (Sect.\,\ref{s-spect}), and the use of spectral energy distributions (SEDs) to determine the spectral type of both components (Sect.\,\ref{s-sed}). Furthermore, if the mass ratio is known from spectroscopy, the SEDs from high-precision broad-band photometry alone can be used to determine their surface gravities with an accuracy up to $\Delta\log{g}$ = 0.05 dex (cgs) and the temperatures with an accuracy as high as 5\%. The resulting surface gravities are independently confirmed by the surface gravities derived from the gravitational redshift (Sect.\,\ref{s-gr}). After which all obtained parameters are summarized (Sect.\,\ref{s-absolutedimensions}). The equivalent widths of iron lines in the spectra can be used to independently confirm the atmospheric parameters of the cool companion (Sect.\,\ref{s-ironlines}). These methods were applied to PG\,1104+243. The cool companion of PG\,1104+243 is presumed to be on the main-sequence, and we will refer to it as the MS component in this paper.  PG\,1104+243 is part of a long-term spectroscopic observing program, and preliminary results of this and seven more systems in this program were presented by \citet{Oestensen11, Oestensen12}.

\section{Spectroscopy}\label{s-spect}
\begin{figure*}[!t]
\centering
\includegraphics{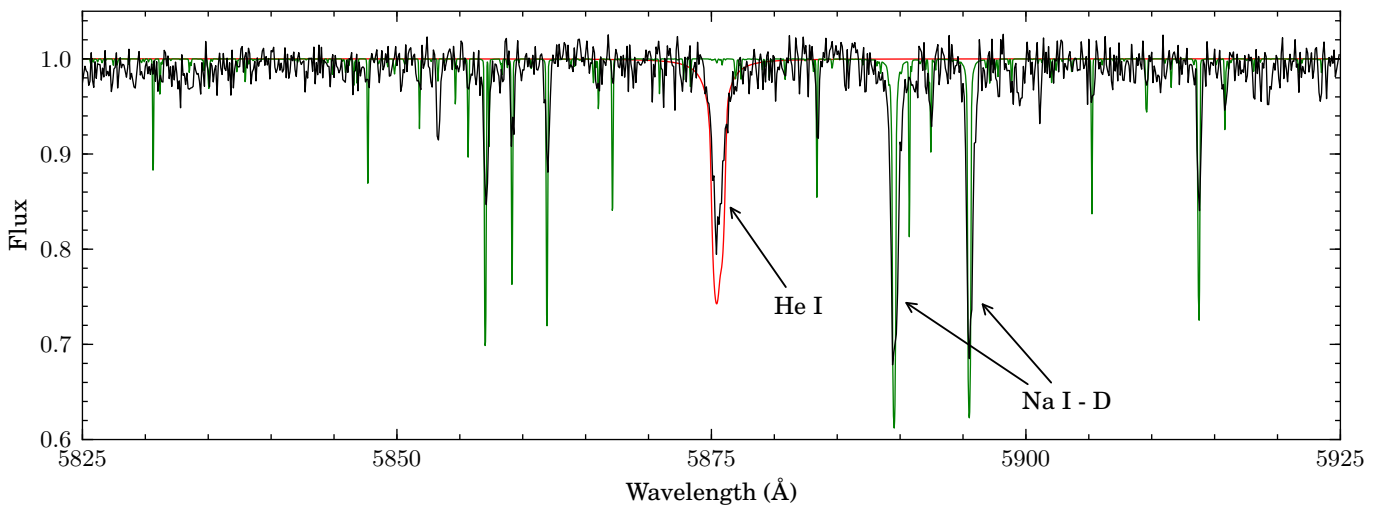}
\caption{A sample normalized spectrum of PG\,1104+243 (black), showing the 5875 \AA\ He\,I line of the sdB component, and several lines as e.g. the Na doublet at 5890-5896 \AA\ of the MS component. Furthermore, a synthetic spectrum of the MS component (green) and sdB component (red) are shown.}\label{fig-spectrum}
\end{figure*}

The high-resolution spectroscopy of PG\,1104+243 was obtained with the HERMES spectrograph (High Efficiency and Resolution Mercator Echelle Spectrograph, R = 85000, 55 orders, 3770-9000 \AA, \citealt{Raskin11}) attached to the 1.2--m Mercator telescope at the Roque de los Muchachos Observatory, La Palma. HERMES is isolated in an over pressurized temperature-controlled enclosure for optimal wavelength stability, and connected to the Mercator telescope with an optical fiber. In total 38 spectra of PG\,1104+243 were obtained between January 2010 and February 2012. The date and exposure time of each spectrum is shown in Table\,\ref{tb-observations}. HERMES was used in high-resolution mode, and Th-Ar-Ne exposures were made at the beginning and end of the night, with the exposure taken closest in time used to calibrate the wavelength scale. The exposure time of the observations was adapted to reach a signal-to-noise ratio (S/N) of 25 in the $V$--band. The HERMES pipeline v3.0 was used for the basic reduction of the spectra, and includes barycentric correction. Part of a sample spectrum taken with HERMES is shown in Fig.\,\ref{fig-spectrum}. 

Furthermore, five intermediate-resolution spectra (R = 4100) were obtained with the Blue Spectrograph at the Multi Mirror Telescope (MMT) in 1996-1997. The Blue Spectrograph was used with the 832/mm grating in 2nd order, covering the wavelength region 4000-4950 \AA.  The data were reduced using standard IRAF tasks. An overview of the spectra is given in Table\,\ref{tb-rv_green}.

\begin{table}
\caption{The observing date (mid-time of exposure), exposure time and the initials of the observer of the PG\,1104+243 spectra observed with HERMES.}
\label{tb-observations}
\centering
\begin{tabular}{llllll}
\hline\hline
BJD		&	Exp.	&	Obs.	&	BJD	&	Exp.	&	Obs.	\\
--2450000	&	s	&		&	--2450000	&	s	&		\\\hline
5204.70741	&	1400	&	K.S.	&	5652.55549	&	1300	&	P.N.	\\
5204.72527	&	1400	&	K.S.	&	5655.60536	&	1400	&	P.N.	\\
5217.67376	&	1500	&	C.W.	&	5658.47913	&	1400	&	P.N.	\\
5217.69173	&	1500	&	C.W.	&	5666.43279	&	1200	&	S.B.	\\
5234.60505	&	900	&	K.E.	&	5666.45153	&	1200	&	S.B.	\\
5234.61608	&	900	&	K.E.	&	5672.53688	&	1500	&	S.B.	\\
5234.62709	&	900	&	K.E.	&	5685.44967	&	1200	&	R.L.	\\
5264.54423	&	2700	&	T.M.	&	5718.44565	&	1600	&	A.T.	\\
5340.42820	&	1200	&	R.L.	&	5914.76439	&	1500	&	P.L.	\\
5351.37760	&	1500	&	S.B.	&	5937.71438	&	1800	&	C.W.	\\
5553.70058	&	1200	&	K.E.	&	5943.63769	&	1800	&	N.G.	\\
5569.69768	&	1200	&	S.B.	&	5953.76375	&	1200	&	J.M.	\\
5579.56553	&	1500	&	P.D.	&	5957.72053	&	1200	&	J.M.	\\
5579.58347	&	1500	&	P.D.	&	5959.59752	&	1200	&	J.M.	\\
5589.75558	&	1200	&	P.L	&	5963.64527	&	3600	&	J.V.	\\
5611.64081	&	1200	&	P.C.	&	5964.53128	&	3500	&	J.V.	\\
5622.61267	&	1200	&	N.C.	&	5964.65639	&	2400	&	J.V.	\\
5639.58778	&	1500	&	N.G.	&	5966.69796	&	2400	&	J.V.	\\
5650.50737	&	1500	&	B.A.	&	5968.61334	&	1800	&	J.V.	\\
\hline
\end{tabular}
\end{table}

To check the wavelength stability of HERMES, 38 different radial velocity standard stars of the IAU were observed over a period of 1481 days, one standard star a night, coinciding with the observing period of PG1104+243. These spectra are cross-correlated with a line mask corresponding to the spectral type of each star to determine the radial velocity. For this cross-correlation (CC), the {\tt hermesVR} method of the HERMES pipeline, which is designed to determine the radial velocity of single stars, is used. This method handles each line separately, starting from a given line-mask. The method uses the extracted spectra, after the cosmic clipping was performed but prior to normalization. It performs a cross-correlation for each line in the mask, and sums the CC functions in each order. The radial velocity is derived by fitting a Gaussian function to the CC function \citep{Raskin11}. To derive the final radial velocity only orders 55$\rightarrow$74 (5966 -- 8920 \AA) are used, as they give the best compromise between maximum S/N for G-K type stars and, after masking the telluric bands, absence of telluric influence \citep{Raskin11}. The resulting standard deviation of these radial velocity measurements is 80 m/s with a non significant shift to the IAU radial velocity standard scale.


\subsection{Radial Velocities}\label{s-radialvelocities}
The determination of the radial velocities of PG\,1104+243 from the HERMES spectra was performed in several steps. First, there were several spectra taken in a short period of time. As the orbital period of PG\,1104+243 was determined at 752 $\pm$ 14 d by \citet{Oestensen12}, spectra that were taken on the same night, or with only a couple of nights in between are summed to increase the S/N. We experimented with several different intervals, and determined that when spectra taken within a five-day interval (corresponding to 0.7\% of the orbital period, or a maximum radial velocity shift of 0.05 km/s) were merged, there was no significant smearing or broadening of the spectral lines. After this merging, 25 spectra with a S/N ranging from 25 to 50 remained.

The MS component has many spectral lines in every Echelle order, most of which are not disturbed by the He lines of the sdB component. Because of this, the CC-method of the HERMES pipeline ({\tt hermesVR}) can be used to determine the radial velocities of the MS component without any loss of accuracy. For the MS component of PG\,1104+243,  a G2-type mask was used. The final errors on the radial velocities take into account the formal errors on the Gaussian fit to the normalized cross-correlation function, the error due to the stability of the wavelength calibration and the error arising due to the used mask. The resulting velocities and their errors are shown in Table\,\ref{tb-rv}.

The determination of the radial velocities of the sdB component is more complicated. Except for the Balmer lines, there are only a few broad He lines visible in the spectrum. After comparison of a synthetic spectrum of a G2-type star with the spectra, it was found that only the 5875.61 \AA\ He\,I blend is not contaminated by lines of the MS component. For this single blend, we had to deploy another method to obtain an accurate radial velocity. As only one line is used, anomalies in the spectra can cause significant errors in the radial velocity determination. To prevent this the region around the He\,I line is cleaned from any remaining cosmic rays by hand. The normalization process is designed to run automatically, making it possible to quickly process multiple lines is many spectra, and consists of two steps. In the first, a spline function is fitted through a 100 \AA\ region centered at the He\,I line to remove the response curve of HERMES, and the main features of the continuum. In the second step, a small 10 \AA\ region around the He\,I line is further normalized by fitting and re-fitting a low degree polynomial through the spectrum. After each iteration, the flux points lower than the polynomial are discarded. This is repeated until the polynomial fits the middle of the continuum. An example of a spectrum normalized using this method is shown in Fig.\,\ref{fig-spectrum}.

These cleaned and normalized spectra can then be cross-correlated to a synthetic sdB spectrum. For this purpose a high-resolution LTE spectrum of $T_{\rm{eff}}$ = 35000 K and $\log{g}$ = 5.50 from the grids of \citet{Heber00} was used. The resolution of this synthetic spectrum is matched to the resolution of the HERMES spectra. Synthetic spectra with different temperatures and surface gravities were tested, to check the temperature and $\log{g}$ dependence of the radial velocities. Changing either of those parameters causes a systemic shift of maximum $0.2\ km/s$ in the resulting radial velocities. The cross-correlation is performed in wavelength space by starting at the expected radial velocity, calculated from the radial velocities of the MS component. To determine the radial velocity, a Gaussian is fitted to the cross-correlation function. The error on the radial velocities is obtained from a Monte-Carlo simulation. This is done by adding Gaussian noise to the spectrum and repeating the cross-correlation with the synthetic spectrum. The level of the Gaussian noise is determined from the noise level in the continuum near the He\,I line. The final error is determined by the standard deviation from the radial velocity results of 1000 simulations, the wavelength stability of HERMES and the dependence on the template. The resulting radial velocities and their errors are shown in Table\,\ref{tb-rv}.

To derive the radial velocities for the MS component from the MMT spectra, the spectra were cross-correlated against two high S/N G0V spectra (HD13974 and HD39587) using the IRAF task {\tt FXCOR}, where the cross-correlation was done only in the wavelength ranges containing absorption lines from the MS companion, ie avoiding the Balmer lines, He\,$\lambda$ 4026, He\,$\lambda$ 4387, He\,$\lambda$ 4471, He\,$\lambda$ 4686, He\,$\lambda$ 4712, and He\,$\lambda$ 4921. The velocities derived from the two G0V stars were averaged to get the velocity of the MS companion for each of the five PG\,1104+243 spectra. The proper scaling factor for the velocity errors output by {\tt FXCOR} was determined using the standard deviation of the velocities obtained from each template relative to the average velocity for each epoch. The resulting radial velocities and errors and are given in Table\,\ref{tb-rv_green}.

\begin{table}
\caption{The radial velocities of both components of PG\,1104+243.}
\label{tb-rv}
\centering
\begin{tabular}{lrrrr}
\hline\hline
\noalign{\smallskip}
	&	\multicolumn{2}{r}{MS component}	&	\multicolumn{2}{r}{sdB component}	\\
BJD	&		RV	&	Error		&		RV	&	Error		\\
-2450000	&	km s$^{-1}$	&	km s$^{-1}$	&	km s$^{-1}$	&	km s$^{-1}$	\\\hline
\noalign{\smallskip}
5204.7163	&	-12.014	&	0.146	&	-19.25	&	1.18  \\
5217.6827	&	-12.036	&	0.150	&	-20.32	&	0.65  \\
5234.6160	&	-11.323	&	0.147	&	-19.66	&	0.23  \\
5264.5442	&	-11.345	&	0.147	&	-20.87	&	0.48  \\
5340.4282	&	-11.916	&	0.146	&	-19.37	&	0.57  \\
5351.3776	&	-12.138	&	0.149	&	-19.19	&	0.56  \\
5553.7005	&	-18.106	&	0.147	&	-9.31	&	0.39  \\
5569.6976	&	-18.679	&	0.150	&	-8.91	&	0.26  \\
5579.5745	&	-19.183	&	0.149	&	-8.54	&	0.36  \\
5589.7555	&	-19.063	&	0.150	&	-8.37	&	0.29  \\
5611.6408	&	-19.607	&	0.149	&	-7.28	&	0.22  \\
5622.6126	&	-19.982	&	0.146	&	-6.74	&	0.54  \\
5639.5877	&	-19.622	&	0.146	&	-6.76	&	0.52  \\
5651.5314	&	-20.112	&	0.144	&	-6.96	&	0.35  \\
5657.0422	&	-19.873	&	0.146	&	-6.25	&	1.02  \\
5666.4421	&	-20.327	&	0.144	&	-6.89	&	1.32  \\
5672.5368	&	-20.259	&	0.147	&	-7.25	&	0.23  \\
5685.4496	&	-20.119	&	0.146	&	-6.81	&	0.48  \\
5718.4456	&	-19.705	&	0.147	&	-6.95	&	0.34  \\
5914.7644	&	-13.405	&	0.147	&	-16.88	&	0.38  \\
5937.7143	&	-12.741	&	0.144	&	-18.19	&	0.22  \\
5943.6376	&	-12.528	&	0.146	&	-18.58	&	0.48  \\
5955.7421	&	-12.361	&	0.147	&	-18.94	&	0.82  \\
5962.5913	&	-12.214	&	0.144	&	-18.84	&	0.71  \\
5966.6559	&	-11.994	&	0.147	&	-19.21	&	0.56  \\
\hline
\end{tabular}
\end{table}

\begin{table}
\caption{The observing date (mid-time of exposure), exposure time and radial velocities of the MS component of PG\,1104+243, determined from the spectra observed with the Blue Spectrograph at the MMT.}
\label{tb-rv_green}
\centering
\begin{tabular}{lrrrr}
\hline\hline
\noalign{\smallskip}
BJD		&	Exp.	&	RV	&	Error		\\
-2450000	&	s	&	km s$^{-1}$	&	km s$^{-1}$\\\hline
\noalign{\smallskip}	
435.0168	&	 300	&	-20.21	&	0.60	\\
436.0515	&	 300	&	-19.21	&	0.63	\\
476.9221	&	 400	&	-18.24	&	0.66	\\
510.8811	&	 120	&	-18.94	&	0.61	\\
836.0087	&	 200	&	-11.94	&	0.58	\\
\hline
\end{tabular}
\end{table}

\subsection{Orbital Parameters}\label{s-orbitalparamters}
The orbital parameters were calculated by fitting a Keplerian orbit through the radial-velocity measurements, while adjusting the period ($P$), time of periastron ($T_0$), eccentricity ($e$), angle of periastron ($\omega$), amplitude ($K$) and systemic velocity ($\gamma$), using the orbital period determined by \citet{Oestensen12} as a first guess for the period. The MS and sdB component were treated separately in this procedure. The radial velocities of both components were weighted according to their errors as $w = 1/\sigma$.
Fitting the orbit resulted in a very low eccentricity ($e < 0.002$). To check if the orbit of the sdB+MS binary is circularized, the \citet{Lucy71} test for circularized orbits was applied:
\begin{equation}
P = \left( \frac{\sum{ (o-c)^2_{\rm{ecc}}} }{\sum{ (o-c)^2_{\rm{circ}}} } \right)^{(n-m)/2},
\end{equation}
where ecc indicates the residuals of an eccentric fit, and circ the residuals of a circular fit, $n$ the total number of observations, and $m = 6$, the number of free parameters in an eccentric fit. If $P  < 0.05$, the orbit is significantly eccentric, if $P \geq 0.05$, the orbit is effectively circular. In the case of PG\,1104+243 we find P = 0.15, indicating a circular orbit. This is expected as the orbit is supposed to be circularized by tidal interactions during the RGB evolution \citep{Zahn77}. In the further determination of the orbital parameters, a circular orbit is assumed.

\begin{figure}
\centering
\includegraphics{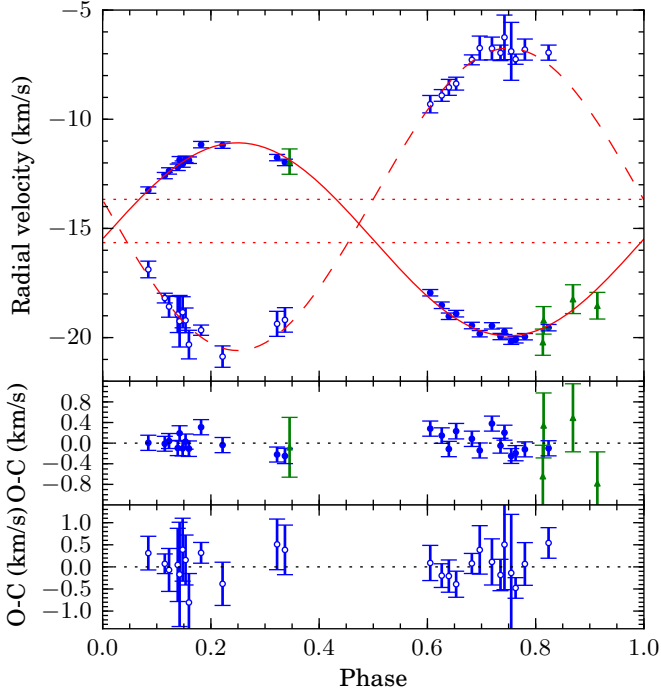}
\caption{Top: spectroscopic orbital solution for PG\,1104+243 (solid line: MS, dashed line: sdB), and the observed radial velocities (blue filled circles: MS HERMES spectra, blue open circles: sdB HERMES spectra, green triangles: MS MMT spectra). The measured system velocity of both components is shown by a dotted line. Middle: residuals of the MS component. Bottom: residuals of the sdB component.}
\label{fig-rvcurves}
\end{figure}

To derive the final orbital parameters, both components were first treated separately to measure the difference in systemic velocity (see Sect.\,\ref{s-gr}). Then that difference was subtracted from the radial velocities of the sdB component, and both components are solved together to obtain their spectroscopic parameters.
The uncertainties on the orbital parameters were determined by using Monte Carlo-simulations. The radial velocities were perturbed based on their errors, and the errors on the parameters were determined by their standard deviation after 5000 iterations. The residuals of the orbital fit to the MS component are larger than the errors on the radial velocities. However, this variability is most likely intrinsic, caused by stellar spots or low amplitude pulsations, and does not affect the derived orbital parameters. The spectroscopic parameters of PG\,1104+243 are shown in Table\,\ref{tb-specparam}. The radial-velocity curves and the best fit solution are plotted in Fig \ref{fig-rvcurves}.


\begin{table}
\centering
\caption{Spectroscopic orbital solution for both the main-sequence (MS) and subdwarf B (sdB) component of PG\,1104+243.}
\label{tb-specparam}
\begin{tabular}{lrr}
\hline\hline
\noalign{\smallskip}
Parameter	&	\multicolumn{1}{c}{MS}	&	\multicolumn{1}{c}{sdB}	\\\hline
\noalign{\smallskip}
$P$ (d)					&	\multicolumn{2}{c}{753 $\pm$ 3}			\\
$T_0$					&	\multicolumn{2}{c}{2450386  $\pm$ 8}		\\
$e$					&	\multicolumn{2}{c}{$<$ 0.002}				\\
$q$ 					&	\multicolumn{2}{c}{0.64$\pm$0.01}		\\
$\gamma$ (km s$^{-1}$)		&	$-$15.63 $\pm$ 0.06	&	$-$13.7 $\pm$ 0.2	\\
$K$  (km s$^{-1}$)		&	4.42 $\pm$ 0.08	&	6.9 $\pm$ 0.2	\\
$a$ $\sin{i}$ ($R_{\odot}$)	&	66 $\pm$ 1	&	103 $\pm$ 1	\\
$M$ $\sin^3{i}$ ($M_{\odot}$)	&	0.069 $\pm$ 0.002	&	0.044 $\pm$ 0.002	\\
$\sigma$			&	0.18			&		0.27		\\
\hline
\end{tabular}
\tablefoot{$a$ denotes the semi-major-axis of the orbit. The quoted errors are the standard deviation from the results of 5000 iterations in a Monte Carlo simulation.}
\end{table}

\section{Spectral energy distribution}\label{s-sed}
To derive the spectral type of the main-sequence and subdwarf component, we fitted the photometric spectral energy distribution (SED) of PG\,1104+243 with model SEDs. With this method we can determine the effective temperature and surface gravity of both components with good accuracy.

\subsection{Photometry}
Photometry of PG\,1104+243 was collected using the subdwarf database\footnote{http://catserver.ing.iac.es/sddb/} \citep{Oestensen06}, which contains a compilation of data on hot subdwarf stars collected from the literature. In total, fifteen photometric observations were found in four different systems: Johnson $B$ and $V$, Cousins $R$ and $I$, Str\"{o}mgren $uvby$, and 2MASS $J$, $H$ and K$_s$. Four observations have uncertainties larger or equal than 0.1 mag and are discarded. The observations used in the SED fitting process are shown in Table\,\ref{tb-photometry}. Accurate photometry at both short and long wavelengths are used to establish the contribution of both the hot sdB component and the cooler MS component.

\begin{table}
\caption{Photometry of PG\,1104+243 collected from the literature.}
\label{tb-photometry}
\centering
\begin{tabular}{lrrrr}
\hline\hline
\noalign{\smallskip}
Band    &   Wavelength  &   Width   &  Magnitude    &   Error   \\
    &   \AA     &   \AA &  mag      &   mag \\\hline
\noalign{\smallskip}
Johnson $B$\tablefootmark{a}    	&   4450	&940    &   11.368  &   0.010   \\
Johnson $V$\tablefootmark{a}    	&   5500 	&880    &   11.295  &   0.013   \\
Cousins $R$\tablefootmark{a}    	&   6500 	&1380   &   11.161  &   0.010   \\
Cousins $I$\tablefootmark{a}    	&   7880 	&1490   &   11.001  &   0.010   \\
Str\"{o}mgren $u$\tablefootmark{b}	&   3460 	&300    &   11.513  &   0.045   \\
Str\"{o}mgren $v$\tablefootmark{b}	&   4100 	&190    &   11.485  &   0.045   \\
Str\"{o}mgren $b$\tablefootmark{b}	&   4670 	&180    &   11.334  &   0.045   \\
Str\"{o}mgren $y$\tablefootmark{b}	&   5480 	&230    &   11.256  &   0.006   \\
2MASS $J$\tablefootmark{c}  		&   12410 	&1500   &   10.768  &   0.026   \\
2MASS $H$\tablefootmark{c}  		&   16500 	&2400   &   10.520  &   0.027   \\
2MASS $K_s$\tablefootmark{c}  		&   21910 	&2500   &   10.510  &   0.023   \\
\hline
\end{tabular}
\tablebib{
\tablefoottext{a}{\citet{Allard94}}
\tablefoottext{b}{\citet{Wesemael92}}
\tablefoottext{c}{\citet{Skrutskie06}}
}
\end{table}

\subsection{SED fitting}
In the SED of PG\,1104+243 in Fig. \ref{fig-sed}, we see both the steep rise in flux towards the shorter wavelengths of the sdB component and the bulk in flux in the red part of the SED of the MS component. To fit a synthetic SED to the observed photometry, Kurucz atmosphere models \citep{Kurucz93} are used for the MS component, and TMAP (T\"{u}bingen NLTE Model-Atmosphere Package, \citealt{Werner03}) atmosphere models for the sdB component. The Kurucz models used in the SED fit have a temperature range from 4000 to 9000 K, and a surface gravity range of $\log{g}$=3.5 dex (cgs) to 5.0 dex (cgs). The TMAP models we have used cover a temperature range from 20000 K to 50000 K, and $\log{g}$ from 4.5 dex (cgs) to 6.5 dex (cgs). They are calculated using an atmospheric mixture of 97 \% hydrogen and 3 \% helium in mass due to the expected He-depletion in the atmosphere of sdB stars. The original TMAP atmosphere models cover a wavelength range of 2500--15000 \AA. To include the 2MASS photometry as well, they are extended with a black body of corresponding temperature to 24000 \AA.

The SEDs are fitted following the grid-based method described in \citet{Degroote11}, but we extended it to include constraints from binarity. In the binary scenario, there are eight free parameters to consider; the effective temperatures ($T_{\rm{eff,MS}}$ and $T_{\rm{eff,sdB}}$), surface gravities ($g_{\rm{MS}}$ and $g_{\rm{sdB}}$) and radii ($R_{\rm{MS}}$ and $R_{\rm{sdB}}$) of both components. The interstellar reddening $E(B-V)$ is naturally presumed equal for both components, and is incorporated using the reddening law of \citet{Fitzpatrick2004}. The models are first corrected for interstellar reddening, and then integrated over the photometric passbands. The distance ($d$) to the system, acts as a global scaling factor. 

The radii of both components are necessary as scale factors for the individual fluxes when combining the atmosphere models of both components to a single SED. Including the distance to the system, the flux of the combined SED is given by:
\begin{equation}
 F_{tot}(\lambda) = \frac{1}{d^2} \left( R_{\rm{MS}}^2 F_{\rm{MS}}(\lambda) + R_{\rm{sdB}}^2 F_{\rm{sdB}}(\lambda) \right),\label{e-sumflux}
\end{equation}
where $F_{\rm{MS}}$ is the flux of the MS model, and $F_{\rm{sdB}}$ is the flux of the sdB model, in a particular passband.
However, if the masses of both components are known, the radii can be derived and removed as free parameters. Assuming that the mass of the sdB component is $M_{\rm{sdB}}$ = 0.47 $M_{\odot}$ as predicted by stellar evolution models (see Sect. \ref{s-intro}), the mass of the MS component can be calculated using the mass ratio derived from the radial velocity curves (see Sect. \ref{s-orbitalparamters}): $M_{\rm{MS}} = 1/q\ M_{\rm{sdB}}$. For each model in the grid, the radii of both components are derived from their surface gravities and masses, according to:
\begin{equation}
 R_i = \sqrt{\frac{G M_i}{g_i}},\: i=\rm{sdB, MS}\label{e-loggR},
\end{equation}
in which $G$ is the gravitational constant. This relation can be used to eliminate the radius dependence in Eq. \ref{e-sumflux}:
\begin{equation}
 F_{tot}(\lambda) = \frac{G\ M_{\rm{MS}}}{d^2\ g_{\rm{MS}}} \left( F_{\rm{MS}}(\lambda) + q \frac{g_{\rm{MS}}}{g_{\rm{sdB}}}\ F_{\rm{sdB}}(\lambda) \right).
\end{equation}
The distance to the system is computed by shifting the combined synthetic model flux ($F_{tot}$) to the photometric observations. As the effect of the mass of the main-sequence component ($M_{\rm{MS}}$) in this equation can be adjusted by the distance, the resulting flux is only dependent on the mass ratio of the components, and not the presumed mass of the sdB component. By using this mass ratio as a limiting factor on the radii of both components, the number of free parameters in the SED fitting process is reduced from eight to six.

To select the best model, the $\chi^2$ value of each fit is calculated using the sum of the squared errors weighted by the uncertainties $(\xi_i)$ on the observations:
\begin{equation}
 \chi^2 = \sum_i{ \frac{(O_i-C_i)^2}{\xi_i^2}},
\end{equation}
with $O_i$ the observed photometry and $C_i$ the calculated model photometry. The expectation value of this distribution is $k = \rm{N}_{\rm{obs}} - \rm{N}_{\rm{free}}$, with N$_{\rm{obs}}$ the number of observations and N$_{\rm{free}}$ the number of free parameters in the fit. In our case, $k=11-6=5$.
Based on this $\chi^2$ statistics the error bars on the photometry can be checked. If the obtained $\chi^2$ is much higher than the expectation value, the errors on the photometry are underestimated. If on the other hand, the obtained $\chi^2$ is much lower than the expectation value, the errors on the photometry are overestimated or we are over fitting the photometry.

If the $\chi^2$-statistics are valid, the uncertainties on the fit parameters can be estimated using the cumulative density function (CDF) of the statistics. To calculate the CDF, the obtained $\chi^2$ values have to be rescaled. The $\chi^2$ of the best fitting model is shifted to the expectation value of the $\chi^2$-distribution, and all other $\chi^2$ values in the grid are scaled relative to the best $\chi^2$. The probability of a model to obtain a certain $\chi^2$ value is given by the CDF, which can be computed using:
\begin{equation}
 F(\chi^2,k) = P\left(\frac{k}{2}, \frac{\chi^2}{2}\right) \label{e-cdf}
\end{equation}
Where P is the regularized $\Gamma$-function. Based on this CDF the uncertainties on the parameters can be calculated based on the distribution of the probabilities.

\subsection{Results}
To fit the SED of PG\,1104+243, first a grid of composite binary spectra is calculated for $\sim$2\,000\,000 points randomly distributed in the $T_{\rm{eff}}$, $\log{g}$ and E(B-V) intervals of both components. These intervals are then adjusted based on the 95\% probability intervals of the first fit, and the fitting process is repeated. To determine final confidence intervals, separate grids of $\sim$1\,000\,000 points which only vary two parameters, while keeping the other parameters fixed at their best fit values, are used. The uncertainties on the parameters are determined based on the 95\% probability intervals. The best fit has a reduced $\chi^2$ of 6.3, and with five degrees of freedom, the $\chi^2$ values in the grid were rescaled by a factor 1.26 which is equivalent to slightly increasing the photometric uncertainties. The SED fit resulted in an effective temperature of 5930 $\pm$ 160 K and 33500 $\pm$ 1200 K for the MS and sdB components respectively, while the surface gravity was determined at respectively 4.29 $\pm$ 0.05 dex and 5.81 $\pm$ 0.05 dex for the MS and sdB component. The reddening of the system was found to be 0.001$^{+0.017}_{-0.001}$ mag. This reddening can be compared to the dust map of \citet{Schlegel98}, which gives an upper bound of E(B-V) = 0.018 for the reddening in the direction of PG\,1104+243. The reddening of  E(B-V) = 0.001$^{+0.017}_{-0.001}$ mag indicates that PG\,1104+243 is located in front of the interstellar molecular clouds. This is confirmed spectroscopically, as there are no sharp interstellar absorption lines (e.g. Ca {\sc ii} K-H, K {\sc i}) visible in the spectra. These parameters indicate that the cool companion is a G-type star.

\begin{figure}
\centering
\includegraphics{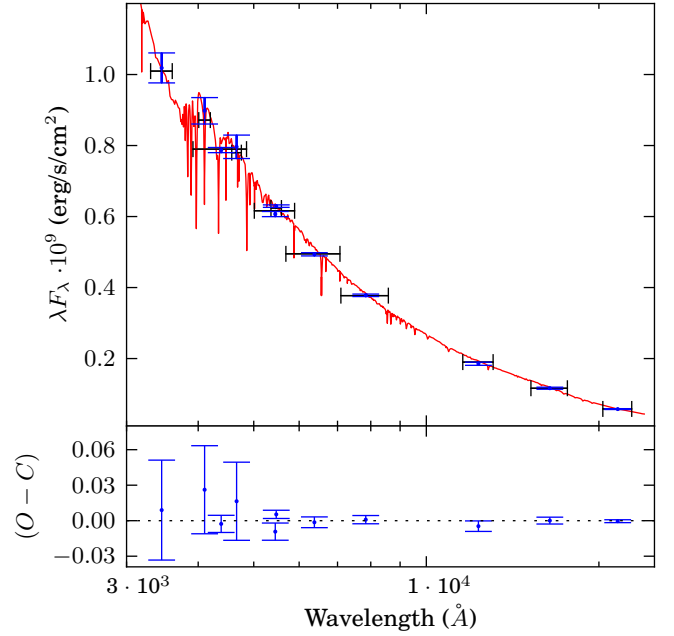}
\caption{Top: The spectral energy distribution of PG\,1104+243. The measurements are given in blue, the integrated synthetic models are shown in black, where a horizontal error bar indicates the width of the passband. The best fitting model is plotted in red. Bottom: The (O-C) value for each synthetic flux point.}
\label{fig-sed}
\end{figure}

\begin{figure*}[!t]
\centering
\includegraphics{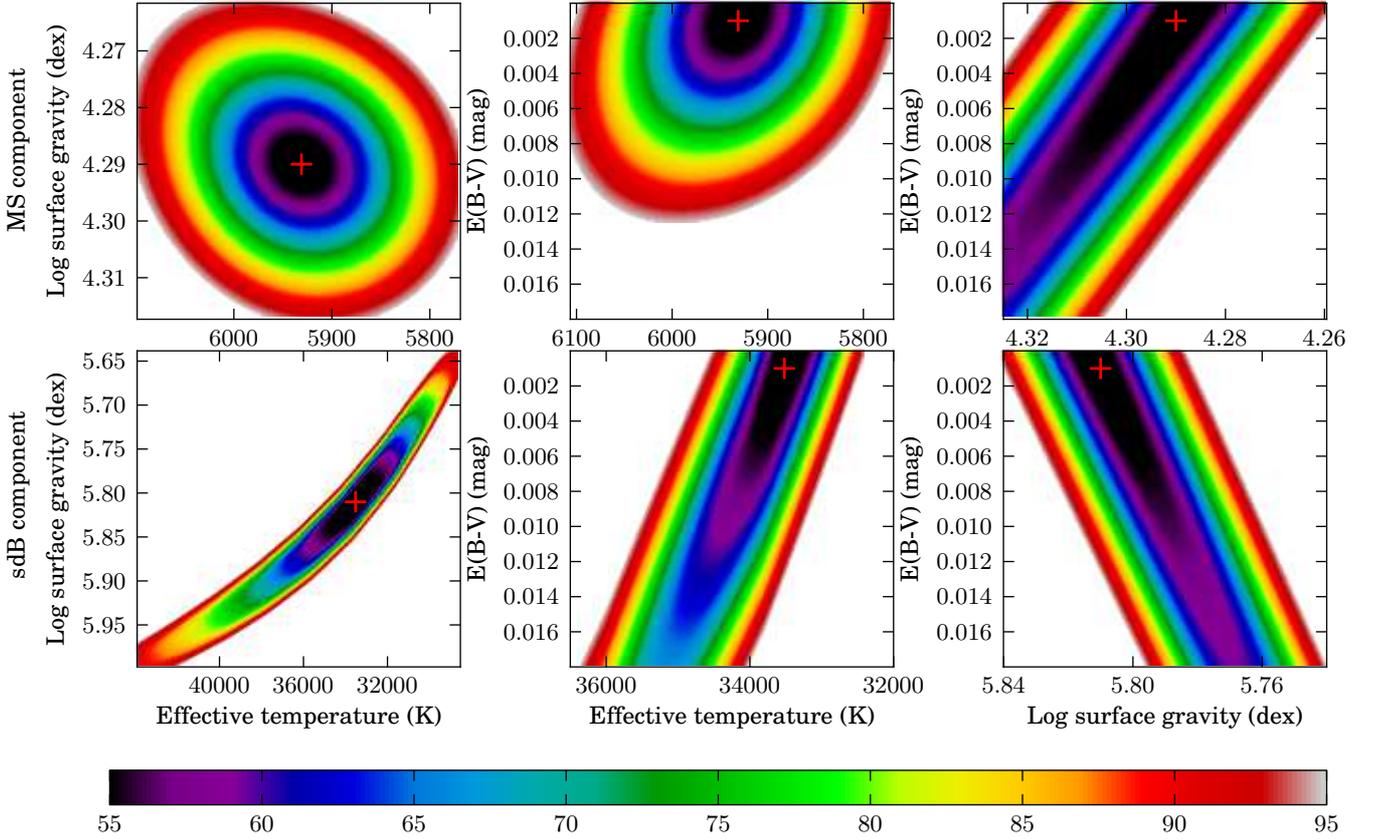}
\caption{The confidence intervals of the SED fit of PG\,1104+243. The upper row show distributions for the main-sequence component, the lower row for the subdwarf component. The best fit results are indicated with a red cross. The different colors show the cumulative density probability connected to the $\chi^2$ statistics given by equation \ref{e-cdf}, with the $\chi^2$ values of the grids rescaled by a factor of 1.26.}\label{fig-grid}%
\end{figure*}

The results of the fit together with the probability intervals are shown in Table\,\ref{tb-sedresults}. The optimal SED fit is plotted in Fig.\,\ref{fig-sed}, while the grids with the probability for each grid point are plotted in Fig.\,\ref{fig-grid}. The probability distribution for the surface gravity and effective temperature of the MS component follows a Gaussian pattern, while the probability distribution for the same parameters of the sdB component is elongated. This latter effect is related to a moderate correlation between $\log{g}$ and $T_{\rm{eff}}$. Thus, the effect on the models caused by an increase in $T_{\rm{eff}}$, can be diminished by increasing $\log{g}$ as well. This correlation is stronger towards shorter wavelengths, so that the main-sequence component, which is mainly visible in the red part of the SED, is less affected. A similar effect is visible in the relation between $E(B-V)$ and $T_{\rm{eff}}$. 

As a check of the results, a SED fit of PG\,1104+243 without any assumptions on the mass of both components is performed as well. In this case, the radii of the components are randomly varied between 0.5 $R_{\odot}$ and 2.0 $R_{\odot}$ for the MS component and between 0.05 $R_{\odot}$ and 0.5 $R_{\odot}$ for the sdB component. The results of this fit are given in the lower half of Table\,\ref{tb-sedresults}. The resulting best fit parameters are very close to the results from the mass ratio-constrained fit, but apart from the uncertainty on the MS effective temperature, the uncertainties on the parameters are much larger. For both the MS and sdB surface gravity, the 95\% probability interval are larger than the range of the models. The effective temperature and surface gravity of both components of PG\,1104+243 correspond with the ionization balance of the iron lines seen in the spectrum of MS component, and the Balmer lines of the sdB component.

\begin{figure}
\centering
\includegraphics{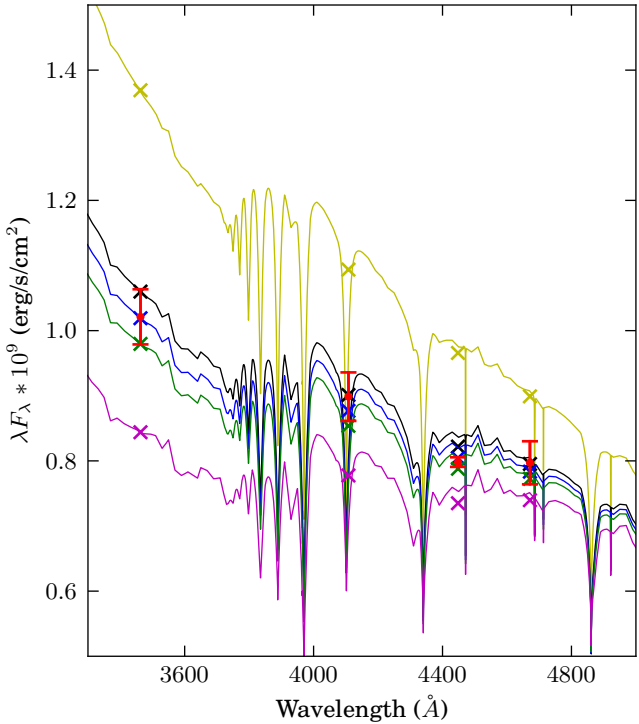}
\caption{SED models calculated for the best fit parameters of PG\,1104+243 as given in Table\,\ref{tb-sedresults}, but with varying surface gravity of the sdB component. From bottom to top, the models have a $\log{g}$ of respectively 6.00, 5.85, 5.81, 5.77 and 5.00 dex (cgs). The crosses show the integrated model photometry. The photometric observations with their error bars are given in red.}
\label{fig-models_sdb}
\end{figure}

The reason for the high accuracy of the mass ratio-constrained fit compared to a fit in which the radii are unconstrained is shown in Fig.\,\ref{fig-models_sdb}. Here five model SEDs are shown, calculated using the best fit parameters from Table\,\ref{tb-sedresults}, but with varying surface gravity, and accordingly adjusted radii, for the sdB component. The best fit model with $\log{g} = 5.81$ dex (cgs) is shown in blue, while models with a $\log{g}$ at the edge of the confidence interval ($\log{g}$ = 5.77 and 5.85 dex) are plotted in black and green. When varying $\log{g}$, the radius of the sdB component changes, while the radius of the MS component remains constant. It is this radius change that causes the large deviation in the models, and allows for an accurate determination of the surface gravity. When only the gravity is changed, and the radii of both components is kept constant, the difference in absolute magnitude in the Str\"{o}mgren-u band is of the order of $10^{-4}$ mag, while when the radius is adapted using the mass, the difference is on the order of $10^{-2}$ mag. Furthermore, because the absolute flux depends on the radius, this method has the advantage that all photometric observations are used to constrain the $\log{g}$ of each component, instead of only the observations in the blue or the red part. This analysis of PG\,1104+243 clearly shows the power of binary SED fitting when the mass ratio of the components is known. 

\begin{table}
\caption{The results of the SED fit, together with the 95\% and 80\% probability intervals. In the upper part (fixed radius) the results when the radii of both components are determined from the $\log{g}$ and mass, are shown. In the lower part (free radius) the results with unconstrained radii are shown. For some parameters the 95\% and 80\% probability intervals could not be determined as they are larger than the range of the models.}
\label{tb-sedresults}
\centering
\begin{tabular}{lrr@{--}lr@{--}l}
\hline\hline
\noalign{\smallskip}
Parameter	&	Best fit	&	\multicolumn{2}{c}{95\%}	&	\multicolumn{2}{c}{80\%} \\\hline
\noalign{\smallskip}
\multicolumn{6}{c}{Fixed radius}\\
\multicolumn{6}{l}{Main sequence component}\\
$T_{\rm{eff}}$ (K)	&	5931	&	5769	&	6095	&	5821	&	6035	\\
$\log{g}$ (dex)	&	4.29	&	4.26	&	4.32	&	4.27	&	4.31	\\
E(B-V) (mag)	&	0.001	&	0.000	&	0.012	&	0.000	&	0.008	\\
\\
\multicolumn{6}{l}{Subdwarf B component}\\
$T_{\rm{eff}}$ (K)	&	33520	&	32400	&	34800	&	32764	&	34290	\\
$\log{g}$ (dex)	&	5.81	&	5.77	&	5.85	&	5.79	&	5.83	\\
E(B-V) (mag)	&	0.001	&	0.000	&	0.013	&	0.000	&	0.008	\\
\\
\multicolumn{6}{c}{Free radius}\\
\multicolumn{6}{l}{Main sequence component}\\
$T_{\rm{eff}}$ (K)	&	5970	&	5730	&	6190	&	5840	&	6130	\\
$\log{g}$ (dex)	&	4.38	&	\multicolumn{2}{c}{/}	&	3.55	&	\,/	\\
E(B-V) (mag)	&	0.001	&	0.000	&	0.100	&	0.000	&	0.090	\\
\\
\multicolumn{6}{l}{Subdwarf B component}\\
$T_{\rm{eff}}$ (K)	&	33810	&	27500	&	45100	&	29000	&	40100	\\
$\log{g}$ (dex)	&	5.90	&	4.70	&	\,/	&	5.25	&	\,/	\\
E(B-V) (mag)	&	0.001	&	0.000	&	0.100	&	0.000	&	0.090	\\
\hline
\end{tabular}
\end{table}

\section{Gravitational Redshift}\label{s-gr}
In an sdB + MS binary the difference in surface gravity between the two components is substantial. The surface gravity of a star gives rise to a frequency shift in the emitted radiation, which is known as the gravitational redshift. General relativity shows that the gravitational redshift as a function of the mass and surface gravity of the star is given by \citep{Einstein16}:
\begin{equation}
 z_g = \frac{1}{c^2} \sqrt{G M g}. \label{e-gr}
\end{equation}
Where $z_g$ is the gravitational redshift, $c$ the speed of light, $G$ the gravitational constant, $M$ the mass, and $g$ the surface gravity. This $z_g$ will effectively change the apparent systemic velocity for the star. In a binary system the difference in surface gravity for both components will be visible as a difference in systemic velocity between both components. As the $z_g$ is proportional to the square root of the surface gravity, this effect is only substantial when there is a large difference in $\log{g}$ between both components, as is the case for compact subdwarfs and main-sequence stars.

The measured difference in systemic velocity can be used to derive an estimate of the surface gravity of the sdB component. Using the mass ratio from the spectroscopic orbit, and the canonical mass of the sdB component, the mass of the MS component can be calculated. Combined with the surface gravity of the MS component from the SED fit, the $z_g$ of the MS component can be calculated. The $z_g$ of the sdB component can be obtained by combining the $z_g$ of the MS component with the measured difference in systemic velocity, and can then be converted to an estimated surface gravity of the sdB component.

Using a canonical value of 0.47 $M_{\odot}$ for the sdB component, and the mass ratio derived in Sect.\,\ref{s-orbitalparamters}, we find a mass of 0.74 $\pm$ 0.07 $M_{\odot}$ for the MS component. The SED fit resulted in a $\log{g}_{\rm{MS}}$ of 4.29 $\pm$ 0.05 dex. Using equation \ref{e-gr}, the gravitational redshift of the MS component is $cz_g$ = 0.46 $\pm$ 0.05 km s$^{-1}$. The measured difference in systemic velocity is 1.97 $\pm$ 0.18 km s$^{-1}$, resulting in a total gravitational redshift of $cz_g$ = 2.43 $\pm$ 0.20 km s$^{-1}$ for the sdB component. This is equivalent to a surface gravity of $\log{g}_{\rm{sdB}}$ = 5.90 $\pm$ 0.08 dex, which is very close to the results of the SED fit.

Other phenomena that can affect the presumed redshift of the MS component, as for example convectional blueshift (see, e.g., \citealt{Dravins82, Takeda12}), are not taken into account as the contribution of such effects is predicted to be around the one-sigma level.

\section{Absolute dimensions}\label{s-absolutedimensions}
Combining the results from the SED fit with the orbital parameters derived from the radial velocities, the absolute dimensions of PG\,1104+243 can be determined. With an assumed mass for the sdB component, the inclination of the system can be derived from the reduced mass determined in Sect. \ref{s-orbitalparamters}. Using this inclination, the semi-major axis of the system can be calculated. The surface gravity of the sdB component as determined by the SED fit and the gravitational redshift correspond within errors. For the final surface gravity, the average of both values is taken. The systemic velocity of PG\,1104+243 is determined by correcting the systemic velocity of the MS component with its gravitational redshift, resulting in $\gamma$ = $-$15.17 $\pm$ 0.07 km s$^{-1}$. The radius of both components is derived from their mass and surface gravity. The absolute dimensions of PG\,1104+243 are summarized in Table\,\ref{tb-absolutedim}

\begin{figure}
\centering
\includegraphics{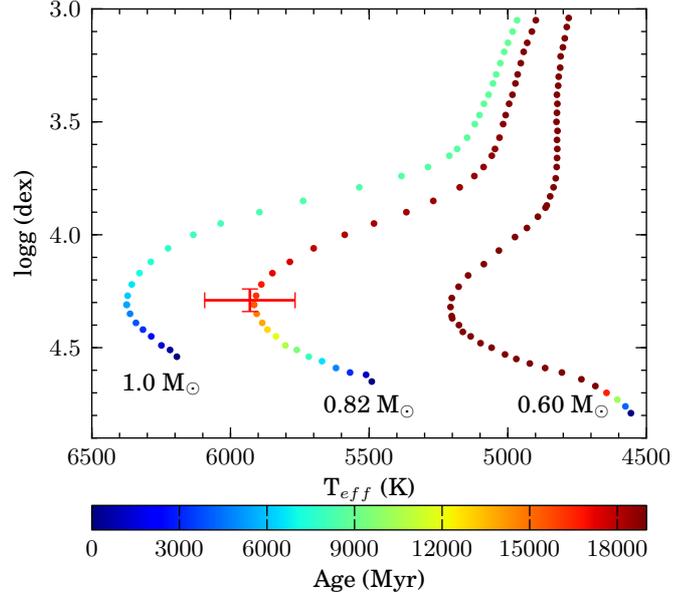}
\caption{Evolutionary tracks calculated by \citet{Bertelli08} with a metalicity of $Z$ = 0.005, for three different masses: 0.60 $M_{\odot}$, 0.82 $M_{\odot}$ and 1.00 $M_{\odot}$. The results obtained for the cool companion of PG\,1104+243 are shown in red.}
\label{fig-tracks}
\end{figure}

For an sdB star of the canonical mass and a surface gravity of $\log{g}$ = 5.85 dex a temperature around 33000 K is expected, consistent with the result of the SED fit and the strength of the \ion{He}{i}/\ion{He}{ii} lines. This high surface gravity and temperature is consistent with the binary population models of \citet{Brown08} that find that low metalicity produces sdB stars located at the high $\log{g}$ end of the extreme horizontal branch. The canonical sdB mass assumption together with the derived mass ratio implies that the MS component must have a sub-solar metalicity. At solar metalicity it should have a radius of 0.8 $R_{\odot}$ and a temperature of $\sim$4800K, which is clearly ruled out by the SED fit. If we drop the canonical sdB mass assumption, and force the MS star to have normal metalicity, both the MS and sdB star must be more massive in order to be consistent with the temperature of the MS component and the spectroscopic mass ratio. From the He-MS models of \citep{Paczynski71}, one finds that a core-He burning model with M $>$ 0.6 $M_{\odot}$ would have Teff $>$ 40000K, which is ruled out by both SED fit and spectroscopy. Recent evolutionary models by \citet{Bertelli08} for main sequence stars with various metalicities, are perfectly consistent with the parameters from the SED fit for an 0.82 $M_{\odot}$ model, provided that the metalicity is assumed to be substantially sub-solar, at around $Z$ = 0.005. Such tracks are shown in Fig.\,\ref{fig-tracks}, where one can see that an 0.82 $M_{\odot}$ star has an expected temperature of around 5900 K, and a surface gravity of $\log{g}$ = 4.30 near the end of the main sequence, corresponding with an age of 15.0 $\pm$ 2.5 Gyr for the MS component. This is similar to the Hubble time and given all uncertainties in the models, indicates that the system is extremely old.

The luminosity of both components can be calculated using $L = 4\pi \sigma R^2 T^4$, resulting in $L_{\rm{MS}}$ = 1.15 $\pm$ 0.13 $L_{\odot}$ and $L_{\rm{sdB}}$ = 22.5 $\pm$ 3.5 $L_{\odot}$. The apparent V magnitudes of both components are obtained directly from the SED fitting procedure and are given by: $V_{0,\rm{MS}}$ = 12.06 $\pm$ 0.05 mag and $V_{0,\rm{sdB}}$ = 12.03 $\pm$ 0.08 mag. The absolute magnitude can be obtained by integrating the best fit model SEDs over the Johnson V band, and scaling the resulting flux to a distance of 10 $pc$, resulting in $M_{V,\rm{MS}}$ = 4.64 $\pm$ 0.05 $mag$ and  $M_{V,\rm{sdB}}$ = 4.61 $\pm$ 0.1 $mag$. The distance to the system can then be calculated from $\log{d} = (m_V − M_V + 5)/5$, which places the system at a distance of $d$ = 305 $\pm$ 10 pc. Obviously, the distance for both components comes out as exactly the same, as this is a fixed condition in the SED fitting process.

The proper motion of PG\,1104+243 as measured by \citet{Hog00} is:
\begin{equation}
 (\mu_{\alpha}, \mu_{\delta}) = (-63.4, -22.1) \pm (1.8, 1.9)\ \rm{mas\ yr^{-1}}
\end{equation}
Using the method of \citet{Johnson87}, these numbers together with the measured value of $\gamma$, can be used to compute the galactic space velocity vector
\begin{equation}
(U,V,W) = (-61.8, -55.3, -52.5) \pm (2.8,2.7,1.3)\ \rm{km\ s^{-1}}
\end{equation}
where U is defined as positive towards the galactic center. Using the values for the local standard of rest from \citet{Dehnen98}, we get
\begin{equation}
(U,V,W)_{\rm{LSR}} = (-51.8, -50.1, -45.3) \pm (2.8,2.8,1.4)\ \rm{km\ s^{-1}}
\end{equation}
Following the selection criteria of \citet{Reddy06}, PG\,1104+243 is bound to the galaxy, and belongs to the thick disk population.

\begin{table}
\centering
\caption{Fundamental properties for both the main-sequence (MS) and subdwarf (sdB) component of PG\,1104+243.}\label{tb-absolutedim}
\begin{tabular}{lrr}
\hline\hline
\noalign{\smallskip}
\multicolumn{3}{l}{Systemic parameters}	\\
$P$ (d)			&	\multicolumn{2}{c}{753 $\pm$ 3}			\\
$T_0$ (HJD)		&	\multicolumn{2}{c}{2450386 $\pm$ 8}		\\
$e$			&	\multicolumn{2}{c}{$<$ 0.002}			\\
$\gamma$ (km s$^{-1}$)	&	\multicolumn{2}{c}{$-$15.17 $\pm$ 0.07}		\\
$q$			&	\multicolumn{2}{c}{0.64 $\pm$ 0.02}			\\
$a$ ($R_{\odot}$)	&	\multicolumn{2}{c}{322 $\pm$ 12}	\\
$i$ $(^o)$		&	\multicolumn{2}{c}{32 $\pm$ 3}			\\
$E(B-V)$		&	\multicolumn{2}{c}{0.001 $\pm$ 0.017}	\\
$d$ (pc)		&	\multicolumn{2}{c}{305 $\pm$ 10}		\\
\noalign{\smallskip}
\multicolumn{3}{l}{Component parameters}	\\
	&	\multicolumn{1}{c}{MS}	&	\multicolumn{1}{c}{sdB}	\\
$K$ (km s$^{-1})$   	&   	4.42   $\pm$   0.08   	&   	6.9   $\pm$   0.3   \\
$M$ ($M_{\odot})$	&	0.74	$\pm$	0.07	&	\multicolumn{1}{c}{0.47 $\pm$ 0.05\tablefootmark{a}}	\\
$\log{g}$ (cgs)		&	4.29	$\pm$	0.05	&	5.85	$\pm$	0.08	\\
$R$ (R$_{\odot})$	&	1.02	$\pm$	0.06	&	0.13	$\pm$	0.02	\\
$T_{\textrm{eff}}$ (K)	&	5930	$\pm$	160	&	33500	$\pm$	1200	\\
$L$ (L$_{\odot}$)       &   1.15    $\pm$  0.13 &   22.5    $\pm$   3.5     \\
$V_0 $			  (mag) &	12.06   $\pm$  0.05 &   12.03   $\pm$   0.08    \\
$M_V$             (mag) &   4.64    $\pm$  0.05 &   4.61    $\pm$   0.10    \\
\hline
\end{tabular}
\tablefoot{\tablefoottext{a}{Assumed value based on evolutionary scenario.}}
\end{table}

\section{Atmospheric parameters from \ion{Fe}{i-ii} lines}\label{s-ironlines}
In this section the equivalent widths of iron lines are used to derive the atmospheric parameters and metalicity of the MS component. This method is independent of the sdB mass assumption used in the SED fit, and is used to confirm the effective temperature and surface gravity obtained with that method. Furthermore, the mass of the sdB component can be derived by using evolutionary tracks to fit the mass of the MS component, and the mass ratio from the radial velocity curves.

Before the EWs can be measured, the continuum contribution of the sdB component must be subtracted from the HERMES spectra, which are then shifted to the zero velocity based on the MS radial velocities and summed. The LTE abundance calculation routine MOOG \citep{Sneden73} was used to determine the atmospheric parameters of the MS component using the EWs of 45 \ion{Fe}{i} and 10 \ion{Fe}{ii} lines in the wavelength range 4400--6850 \AA, excluding blended lines, and wavelength ranges contaminated by Balmer lines, \ion{He}{i} or \ion{He}{ii} lines from the sdB component. The atmospheric parameter determination also provides the metalicity of the MS component. The atomic data for the iron lines was taken from the VALD line lists \citep{Kupka99}, and the used oscillator strengths were laboratory values. The EWs are measured via integration and abundances are computed by an iterative process in MOOG in which for a given abundance, the theoretical EWs of single lines are computed and matched to the observed EWs. The effective temperature is derived by the assumption that the abundance of individual Fe I lines is independent of lower excitation potential, only Fe I lines are used because they cover a large range in lower excitation potential, resulting in $T_{\rm{eff}}$ = 6000 $\pm$ 250 K. A surface gravity of $\log{g}$ = 4.5 $\pm$ 0.5 dex is derived by assuming ionization balance between the iron abundance of individual \ion{Fe}{i} and \ion{Fe}{ii} lines. Assuming an independence between the iron abundance and the reduced EW, a microturbulence velocity of 2.0 $\pm$ 0.5 km/s can be derived. The final metalicity is based on the average abundance of all used lines, and result in a metalicity of [Fe/H] = -0.58 $\pm$ 0.11 dex, corresponding to $Z$ = 0.005 $\pm$ 0.001 using the conversion of \citet{Bertelli94}. The error on the metalicity is the line-to-line scatter for the Fe abundances for the derived atmospheric parameters. The used Fe I lines are fitted by creating synthetic spectra in MOOG using the derived atmospheric parameters, which confirm the obtained metalicity. A detailed description of the method used to determine the atmospheric parameters can be found in \citet[Sect. 3]{DeSmedt12}.
These atmospheric parameters correspond very well with the effective temperature and surface gravity derived from the SED fitting process.

The atmospheric parameters derived from the spectral analysis can be used to get a mass estimate for both the MS and sdB component independently from the SED fitting process, by using evolutionary models. Solar scaled evolutionary tracks of \citet{Bertelli08} with a metalicity of $Z$ = 0.005 are compared to the derived surface gravity ($\log{g}$ = 4.5 $\pm$ 0.5 dex) and effective temperature ($T_{\rm{eff}}$ = 6000 $\pm$ 250 K). The best fitting model has a mass of $M_{\mathrm{MS}}$ = 0.86 $\pm$ 0.15 $M_{\odot}$, with as corresponding age 12 $\pm$ 5 Gyr. Using the mass ratio from the radial velocity curves, a mass of $M_{\rm{sdB}}$ = 0.54 $\pm$ 0.10 $M_{\odot}$ is obtained for the sdB component. This corresponds within error to the canonical value of $M_{\mathrm{sdB}}$ = 0.47 $\pm$ 0.05 $M_{\odot}$.

Using the newly obtained mass of the sdB component to derive its surface gravity from the gravitational redshift we find $\log{g}_{\rm{sdB}}$ = 5.84 $\pm$ 0.08 dex. Furthermore, the radius, luminosity and magnitudes in the V band can be derived using the methods described in Sect.\,\ref{s-absolutedimensions}, the results are shown in Table \ref{tb-absolutedim2}. Most of the derived parameters are compatible with those derived in Sect.\,\ref{s-absolutedimensions}.

\begin{table}
\centering
\caption{Atmospheric parameters obtained based on the EW of iron lines, and derived properties. See Sect.\,\ref{s-ironlines} for further explanation.}\label{tb-absolutedim2}
\begin{tabular}{lrr}
\hline\hline
\noalign{\smallskip}
 	&	\multicolumn{1}{c}{MS}	&	\multicolumn{1}{c}{sdB}	\\\hline
\noalign{\smallskip}
$M$ (M$_{\odot})$	&	0.86	$\pm$	0.15	&	0.54	$\pm$	0.10	\\
$\log{g}$ (cgs)		&	4.50	$\pm$	0.50	&	5.84	$\pm$	0.08	\\
$R$ (R$_{\odot})$	&	0.86	$\pm$	0.07	&	0.15	$\pm$	0.05	\\
$T_{\textrm{eff}}$ (K)	&	6000	$\pm$	250	&	\multicolumn{1}{c}{/}	\\
$L$ (L$_{\odot}$)       &	0.87    $\pm$	0.20	&	\multicolumn{1}{c}{/}	\\
$V_0 $	(mag) 		&	12.30   $\pm$	0.15	&	\multicolumn{1}{c}{/}	\\
$M_V$   (mag) 		&	4.95    $\pm$	0.15	&	\multicolumn{1}{c}{/}	\\
$d$	(pc)		&	296	$\pm$	12	&	\multicolumn{1}{c}{/}	\\
\hline
\end{tabular}
\end{table}

\section{Discussion and Conclusions}

From an analysis of literature photometry and observed spectra, detailed astrophysical parameters of PG\,1104+243 have been established. The surface gravity determined from the binary SED fitting method agrees very well with the surface gravity determined from the observed gravitational redshift. Furthermore, the long time-base observations made it possible to accurately establish the orbital period at 753 $\pm$ 3 d. 

The sdB component is consistent with a canonical post-core-helium-flash model with a mass of $\sim$0.47 $M_{\odot}$, formed through stable Roche lobe overflow. This is supported by the essentially circular orbit as stable mass transfer is the only known process that can circularize a long-period binary system.

When comparing our results with the distribution predicted by BPS models, we find that the orbital period of PG\,1104+243 is higher than the most likely outcome at just over 100 days resulting from \citet[Fig. 21]{Han03}. However, the period is in agreement with the observed estimate of three to four years noted by \citet{Green01}, and with BPS models computed within the $\gamma$-formalism \citep{Nelemans10}. Our results therefore provide strong observational constraints for which BPS models are viable. \citet{Clausen12} published several limitations on the limiting mass ratio for dynamically stable mass transfer during the RLOF phase, the way that mass is lost to space and the transfer of orbital energy to the envelope during a common envelope phase, based on the observed period distribution of sdB + MS binaries. They state that if long-period (P$_{\rm{orb}}$ $>$ 75 d) sdB+MS binaries exist, mass transfer on the red giant branch (RGB) is stable, and if sdB + MS binaries with periods over 250 d exists, mass that leaves the system carries away a specific angular momentum proportional to that of the donor. In both cases the efficiency at which orbital energy is transferred to the envelope during the common envelope phase is below 75 \%. Furthermore, in several of their BPS models the resulting distributions for sdB + MS binaries would include periods up to 500 days \citep[Fig. 16]{Clausen12}.
However, this formation channel cannot explain the low mass of the cool companion. In this formation channel, the cool companion would be expected to have a mass $\sim 1 M_{\odot}$. 

The low mass of the MS component (0.735 $\pm$ 0.07 $M_{\odot}$) indicates that during the stable Roche--Lobe overflow phase only a small amount of the mass was accreted, and the major part must have been lost to infinity. The high evolutionary age of the MS component ($\sim$13000 Myr) is further evidence of this. If the companion had accreted a significant fraction of the original envelope of the sdB component, its progenitor should have had a much lower mass as the current 0.735 $\pm$ 0.07 $M_{\odot}$. In this case, the MS component should be practically un-evolved, as e.g. the 0.6 $M_{\odot}$ track of Fig.\,\ref{fig-tracks} which hardly evolves at all in a Hubble time. As the time since the mass transfer ended cannot be longer than the core-helium burning lifetime of only ~100 Myr, no significant evolution of the MS component can have taken place since then.

\citet{Deca12} published a study of the long-period sdB+K system PG\,1018--047, which shares many similar properties with PG\,1104+243. It has a period of 760 $\pm$ 6 d, a mass ratio of $M_{\rm{sdB}} / M_{\rm{K}}$ = 0.63 $\pm$ 0.11, and they find an effective temperature of 30500 $\pm$ 200 K and $\log{g}$ of 5.50 $\pm$ 0.02 dex for the sdB component. However, while the cool companion in PG\,1104+243 contributes $\sim$52\,\% of the light in the $V$-band, in PG\,1018--047 it only contributes $\sim$6\,\%, and is photometrically and spectroscopically consistent with a mid-K type star. \citet{Deca12} speculate that the orbit of PG\,1018-047 may be quite eccentric, and that it can have formed through the merger scenario of \citet{Clausen11}, meaning that the K-star was not involved in the evolution of the sdB star. The extremely low eccentricity of PG\,1104+243, makes such a scenario very unlikely, and we conclude that it is the first sdB+MS system that shows consistent evidence for being formed through post-Roche-lobe overflow.

Thus, we have used high-resolution spectroscopy to solve the orbits of both components in an sdB+MS binary system, and have for the first time derived accurate and consistent physical parameters for both components. Furthermore, the accurate radial velocities  allowed the first measurement of gravitational redshift in an sdB binary. PG\,1104+243 is part of an ongoing long-term observing program of sdB+MS binaries with HERMES at Mercator, and an analysis of the complete sample will make it possible to refine several essential parameters in the current formation channels.

\begin{acknowledgements}
  We thank the referee for his useful suggestions. 
  Many thanks to N. Cox, N. Gorlova and C. Waelkens for their help with obtaining spectra at the Mercator Telescope.
  Based on observations made with the Mercator Telescope, operated on the island of La Palma by the Flemish Community, at the Spanish Observatorio del Roque de los Muchachos of the Instituto de Astrofísica de Canarias.
  Based on observations obtained with the HERMES spectrograph, which is supported by the Fund for Scientific Research of Flanders (FWO), Belgium , the Research Council of K.U.Leuven, Belgium, the Fonds National Recherches Scientific (FNRS), Belgium, the Royal Observatory of Belgium, the Observatoire de Genève, Switzerland and the Thüringer Landessternwarte Tautenburg, Germany.
  Some of the observations reported in this paper were obtained at the MMT Observatory, a facility operated jointly by the University of Arizona and the Smithsonian Institution.
  The following Internet-based resources were used in research for this paper: the NASA Astrophysics Data System; the SIMBAD database and the VizieR service operated by CDS, Strasbourg, France; the ar$\chi$ive scientific paper preprint service operated by Cornell University.
  This publication makes use of data products from the Two Micron All Sky Survey, which is a joint project of the University of Massachusetts and the Infrared Processing and Analysis Center/California Institute of Technology, funded by the National Aeronautics and Space Administration and the National Science Foundation.
  The research leading to these results has received funding from the European Research Council under the European Community's Seventh Framework Programme (FP7/2007--2013)/ERC grant agreement N$^{\underline{\mathrm o}}$\,227224 ({\sc prosperity}), as well as from the Research Council of K.U.Leuven grant agreement GOA/2008/04, the German Aerospace Center (DLR) under grant agreement 05OR0806 and the Deutsche Forschungsgemeinschaft under grant agreement WE1312/41-1. P. Neyskens is Boursier FRIA, funded by the Fonds National Recherches Scientific.
\end{acknowledgements}

\bibliographystyle{aa}
\bibliography{references}

\end{document}